\documentclass[12pt]{iopart}
\usepackage{iopams} 
\usepackage[latin1]{inputenc}
\usepackage{graphicx}
\usepackage{url}

\def\Re{{\rm Re\mit}}
\def\Im{{\rm Im\mit}}

\begin{document} 
 
\title[Resonance energy transfer -- graphene waveguides -- surface plasmon polaritons]
{Spatial modulation of the electromagnetic energy transfer by excitation of graphene waveguide surface plasmons} 
\author{Julieta Olivo$^1$} 
\address{$^1$Universidad Nacional de R\'io Cuarto, Dto. F\'isica, Facultad de Ciencias Exactas Fisicoqu\'imicas y Naturales, R\'io Cuarto (5800), C\'ordoba, Argentina}
\author{Carlos J. Zapata--Rodr\'iguez$^2$} 
\address{$^2$Department of Optics and Optometry and Vision Science, University of Valencia, Dr. Moliner 50, Burjassot 46100, Spain}
\author{Mauro Cuevas$^3$}
\address{$^3$Consejo Nacional de Investigaciones Cient\'ificas y T\'ecnicas (CONICET) and Facultad de Ingenier\'ia y Tecnolog\'ia Inform\'atica, Universidad de Belgrano,
Villanueva 1324,  Buenos Aires, Argentina}
\ead{cuevas@df.uba.ar}

\begin{abstract} 
We theoretically study the electromagnetic energy transfer between donor and acceptor molecules near a graphene waveguide. The surface plasmons (SPs) supported by the structure provide decay channels which lead to an improvement in the energy transfer rate  when the donor and acceptor are localized on the same side or even on opposite sides of the waveguide. The modification of the energy transfer rate compared to its value in absence of the waveguide are calculated by deforming the integration path into a suitable path in the complex plane.   Our results show that this modification is dramatically enhanced when  the symmetric and antisymmetric SPs are excited.   
Notable effects on the spatial dependence of the energy transfer due to the coherent interference between these  SP channels, which can be tuned by chemical potential variations, are highlighted and discussed in terms of SP propagation characteristics. 
\end{abstract} 



\maketitle

\section{Introduction} 

The electromagnetic resonance energy transfer (RET) between pairs of quantum emitters, such as atoms, molecules or quantum dots, can be largely altered by excitation of electromagnetic eigenmodes.   By coupling these emitters to a designed wave guiding or surface plasmon (SP) mode environments, a significant directional  control and RET enhancement has been achieved \cite{barnes,arruda1,pra84,shahbazyan,maier_nature,wubs} allowing energy transfer over distances larger than the F\"orster energy transfer range \cite{forster}. New avenues have emerged with the advent of graphene, a monolayer of carbon atoms arranged in a hexagonal lattice, which has motivated an extensive wealth of theoretical analysis concerning the shift of 
molecular interplay control via metal SPs, currently working in the visible range, to the mid--infrarred and terahertz frequencies \cite{superradiance,grafeno3,QI}.

 Key representatives of SPs on doped graphene are those with $p$ polarization,  existing below a critical frequency depending of the chemical potential on graphene, which offer high confinement,  relatively low loss and good tunability of its spectrum through electrical or chemical modification of the carrier density \cite{jablan,rana}. 

Because of their fundamental properties as well as their potential applications, the knowledge about the interaction between 
graphene and electromagnetic radiation via SP mechanism 
is a topic of continuously increasing interest, opening the route towards a wide spectrum of studies ranging from photonic devices capable to achieve invisibility \cite{Naserpour,Alu} and THz antennas \cite{filter,gomez_diaz,jornet,cuevas6}  to graphene SP structures capable to control the spontaneous emission as well as the RET  between  quantum emitters \cite{marocico,arruda2,LSP,grafeno2,cuevas3,cuevas3bis,cuevas4}. 

In this paper we study the energy transfer process between a donor and acceptor placed in close proximity to a waveguide formed by two parallel graphene sheets with an insulator spacer layer. Particular interest is paid to the role of the SPs of the structure in modifying the energy transfer rate with respect to the rate
in absence of the waveguide. This issue has been addressed for a single graphene sheet \cite{marocico,biehs,shahbazyan2}, where the main results have shown a broadband and long--range energy transfer enhanced beyond 
four even reaching six orders of magnitude  relative to its value without graphene sheet.  
Even though we expect similar features when the SP modes of the graphene waveguide are excited, one of the interesting differences with a single graphene monolayer structure, which motivate the present work, is that the graphene waveguide under study has two conducting interfaces, each of which may carry SP modes, and the fields of these modes can overlap through the gap dielectric layer, leading SPs into separated branches. Depending on the molecular location and orientation, these modes  can be excited with distinct strength and, as a consequence the coherent interference between SP branches leads  to a strong modulation on the energy transfer rate.
Although an oscillatory behavior due to interference between different modes excited into the intermolecular spacing has been reported for metallic structures, like wires \cite{wire} and  waveguides \cite{pra84}, here the separation between the SP branches may be large enough \cite{cuevas0} to produce a high frequency spatial  modulation and,   such separation and thus the modulation period can be  controlled with the help of a gate voltage on the graphene sheets. 

This paper is organized as follows. In section 2, 
we develop an analytical method based on the separation of variables approach and obtain a solution for the electromagnetic field that is emitted by an oscillating dipole and is scattered by a graphene waveguide. By virtue of the translational invariance of the system along a  plane parallel to graphene sheets, we reduce the solution of the original vectorial problem to the treatment of two scalar problems corresponding to the basic modes of polarization $p$ (magnetic field parallel to the waveguide) and  $s$ (electric field parallel  to the waveguide) for which we derive integral expressions for the scattered electric fields. We then include a second oscillating dipole  and deal with the problem of the coupled system, providing an expression to calculate the energy transfer between the existing two dipoles. By using contour integration in the complex plane, we have developed two methods to perform the field integration: i)  the residues method, which enables  to calculate the contribution of each one of the SP branches to the energy transfer rate and, ii) the direct integration method, in which a suitable path deformation in the complex plane enables an accurately evaluation of the integral. In section 3 
we present numerical results obtained under different dipole moment configurations. Concluding remarks are provided in Section 4. 
The Gaussian system of units is used and an $\mbox{exp}(-i\, \omega\, t)$ time--dependence is implicit throughout the paper, where $\omega$ is the angular frequency, $t$ is the time coordinate, and $i=\sqrt{-1}$. The symbols Re and Im are used for denoting the real and imaginary parts of a complex quantity, respectively.

\section{Theory}
\label{teoria}

\subsection{Electromagnetic energy transfer between two dipole emitters}

We consider the energy transfer rate between a donor D and an acceptor A electric dipoles placed close to a planar graphene  waveguide, as illustrated in Figure 1. In accordance with Poynting theorem the time--average power transferred from the dipole $p_D$  to dipole $p_A$ can be calculated by means of
\begin{eqnarray}\label{ET}
P_{ET}= -\frac{1}{2}\int_{V_A} \Re \left\{\textbf{j}_A^{*}(\textbf{x}) \cdot 
\textbf{E}_D(\textbf{x})\right\} d^3x,
\end{eqnarray}
where $V_A$ encloses the acceptor dipole $p_A$,  $\textbf{j}_A$ represents the source density current associated with the dipole $p_A$ and $\textbf{E}_D$ is the electric field generated by the donor  dipole $p_D$ \cite{novotny}.
\begin{figure}
\centering
\resizebox{0.6\textwidth}{!}
{\includegraphics{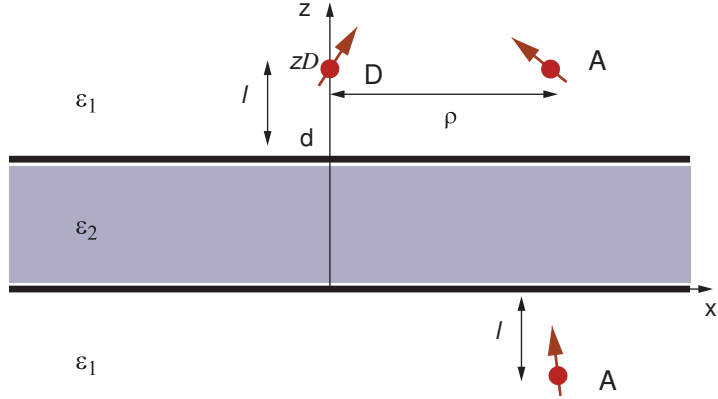}}
\caption{\label{fig:epsart} Schematic illustration of the system. 
Two graphene encapsulation layers of surface conductivity $\sigma$ are deposited on an embedded dielectric slab
 at $z = 0$ and $z = d$. The donor D is located at $\textbf{x} = z_D \hat{z}$ ($z_D=d+l$) and acceptor A is located at $\textbf{x}=\rho \hat{\rho}+z_D\hat{z}$ (above the waveguide) or $\textbf{x}=\rho \hat{\rho}-l\hat{z}$ (below the waveguide).
}\label{sistema}
\end{figure}

In this approximation, the current density $\textbf{j}_A(\textbf{x})=-i \omega \textbf{p}_A \, \delta(\textbf{x}-\textbf{x}_A)$   and, taking into account that the dipole moment $p_A$ is the induced dipole by the electric field  $\textbf{E}_D$ generated by $p_D$, in the linear regime, Eq. (\ref{ET}) can be written as \cite{novotny}
\begin{eqnarray}\label{ET2}
P_{ET}=\frac{\omega}{2} \Im\left\{\alpha_A\right\}|\hat{n}_A\cdot \textbf{E}_D(\textbf{x}_A)|^2,
\end{eqnarray}
where $\alpha_A$ is the polarizability of the aceptor and $\hat{n}_A$ is a unit vector along the induced polarization of the acceptor (whose direction is supposedly to be fixed). The energy transfer rate normalized with respect to the power emitted in absence of the waveguide, $P_0=\frac{\omega p^2_D k_1^3}{3 \varepsilon_1}$, is written as 
\begin{eqnarray}\label{ET3}
\frac{P_{ET}}{P_0}=\Im\left\{\alpha_A\right\} T(\omega),
\end{eqnarray}
where 
\begin{eqnarray}\label{T}
T(\omega)=\frac{3 \varepsilon_1}{2 k_1^3 p_D^2}|\hat{n}_A\cdot \textbf{E}_D(\textbf{x}_A)|^2
\end{eqnarray}
is the energy transfer function. The energy transfer  in the presence of the graphene waveguide normalized to that in an unbounded medium 1 (without graphene waveguide), is given by
\begin{eqnarray}\label{ETn}
F_{ET}=\frac{T(\omega)}{T_0(\omega)}, 
\end{eqnarray}
with 
\begin{eqnarray}\label{T}
T_0(\omega)=\frac{3 \varepsilon_1}{2 k_1^3 p_D^2}|\hat{n}_A\cdot \textbf{E}_0(\textbf{x}_A)|^2,
\end{eqnarray}
where $\textbf{E}_0(\textbf{x}_A)$ is the field generated by the donor at the acceptor position in the absence of the graphene waveguide.

The expression of the electric field $\textbf{E}_0(\textbf{x})$ due to an electric dipole in an unbounded medium is given in  \ref{ap1}, and a detailed sketch to obtain integral expressions for the electric field $\textbf{E}_D(\textbf{x})$  in the presence of a graphene waveguide is presented in \ref{ap2}.

\subsection{Case $\textbf{p}_D=p\hat{z}$}
Firstly, we consider the case in which the donor dipole moment is oriented in the $\hat{z}$ direction. From Eq. (\ref{pym}), it can be seen that $d_p=p$ and $d_s=0$ and, as a consequence, $A^{(1)}_s=B^{(3)}_s=0$.     Even though Eqs. (\ref{campoE1}) and (\ref{campoE3}) give the electric field outside the waveguide straightforwardly, it is convenient to carry out their transformation to polar coordinates,
\begin{eqnarray}\label{polares}
\alpha=k_{||} \cos \phi_k,\\
\beta=k_{||} \sin \phi_k.
\end{eqnarray}
Taking into account the integral definition of the $0$th order of Bessel function
\begin{eqnarray}\label{polares}
J_0(x)=\frac{1}{2\pi}\int_0^{2\pi} du e^{ix \cos u}, 
\end{eqnarray}
and the recurrence relation 
\begin{eqnarray}\label{polares}
J'_m(x)=\frac{1}{2}(J_{m-1}(x)-J_{m+1}(x)), 
\end{eqnarray}
 Eqs. (\ref{campoE1}) and (\ref{campoE3}) which give the scattered electric field expressions in medium 1 (above the waveguide) and in medium 2 (below the waveguide) can be rewritten as
\begin{eqnarray}\label{campoE1b}
\textbf{E}^{(1)}(\textbf{x})|_{scatt}=\nonumber\\
\frac{i k_0^2}{k_1^2}\int_{0}^{+\infty}  \left[
\gamma^{(1)} \frac{J_1(k_{||}\rho)}{i} \hat{\rho}
+k_{||}J_0(k_{||}\rho)\hat{z} 
 \right] A^{(1)}_p   e^{i \gamma^{(1)}z} k_{||}^2 dk_{||}
 ,    
\end{eqnarray}
and 
\begin{eqnarray}\label{campoE3b}
\textbf{E}^{(3)}(\textbf{x})=\nonumber \\
\frac{i k_0^2}{k_1^2}\int_{0}^{+\infty}  \left[
\gamma^{(1)} \frac{J_1(k_{||}\rho)}{i} \hat{\rho}
+k_{||}J_0(k_{||}\rho)\hat{z} 
 \right] B^{(3)}_p   e^{-i \gamma^{(1)}z} k_{||}^2 dk_{||}
 . 
\end{eqnarray}
An integral of this kind represents a challenge due to the uncomfortable behavior of the integrand. To avoid this difficulty we have developed two methods to perform such integration. 
The first method requires the application of the residues theorem to  extract each pole contribution of the integrals (\ref{campoE1b}) and (\ref{campoE3b}).  The second method is carried out by using the Cauchy's integral theorem in order to deform the path of integration into a suitable path in  the complex $k_{||}$ plane. Then we use a Gauss--Legendre quadrature to evaluate the integrals along the deformed path.

\subsubsection{Calculation of the graphene eigenmodes contribution} \label{modos grafeno}

The integration path in Eqs. (\ref{campoE1b}) and (\ref{campoE3b}) is set along the real and positive $k_{||}$ axis, so that the integral will be strongly  affected by singularities that are close to that axis. Pole singularities, i.e., zeroes of the denominator  in $A_p^{(1)}$ and $B_p^{(3)}$ coefficients, occur at generally complex locations ($k_{||}$ is a complex magnitude) and they represent the propagation constant of the eigenmodes supported by the graphene waveguide, like waveguide (WG) or surface plasmon (SP) modes. The WG modes refer to modes which are evanescent waves in the two semi infinite regions (regions 1 and 3) and standing waves in the insulator spacer layer (region 2), and SPs refer to modes which propagate along the waveguide with their electric and magnetic fields decaying exponentially away from the graphene sheets in all three regions. 
%
%
These eigenmodes (especially SP modes)  provide new decay channels for the electromagnetic energy transfer 
between single emitters placed close to the waveguide. For this reason,  close attention must be paid to the calculation of  their contribution  to the field integrals (\ref{campoE1b}) and (\ref{campoE3b}) as well as to the resonance energy transfer (\ref{ETn}).
%
%
The integration method described in the present subsection makes use of the symmetry properties of the Bessel and Hankel functions \cite{abramowitz} given by
\begin{eqnarray}\label{bessel}
J_n(x)=\frac{1}{2\pi}[H_n^{(1)}(x)+H_n^{(2)}(x)],\nonumber \\
H_n^{(1)}(x e^{i \pi})=-e^{-i n \pi}H_n^{(2)}(x).
\end{eqnarray}
%
Application of the symmetry properties (\ref{bessel}) to the functions that form our integrands in (\ref{campoE1b}) and (\ref{campoE3b}), results in the following identities:
\begin{eqnarray}\label{paridad}
\int_0^{+\infty} f_{odd}(x) J_0(x) dx=\frac{1}{2}\int_{-\infty}^{+\infty} f_{odd}(x) H_0^{(1)}(x) dx ,\nonumber \\
\int_0^{+\infty} f_{even}(x) J_1(x) dx=\frac{1}{2}\int_{-\infty}^{+\infty} f_{even}(x) H_1^{(1)}(x) dx ,
\end{eqnarray}
with
\begin{eqnarray}\label{fimparA}
f_{odd}(x)= \frac{i}{\varepsilon_1} O \,  
k_{||}^3, \nonumber \\
f_{even}(x)= \frac{1}{\varepsilon_1} O \, 
\gamma^{(1)} k_{||}^2
\end{eqnarray}
where $O=A^{(1)}_p\,e^{i \gamma^{(1)}z}$ for the field components in Eq. (\ref{campoE1b}) or $O=B^{(3)}_p\,e^{-i \gamma^{(1)}z}$ for the field components in Eq. (\ref{campoE3b}). Therefore, Eqs. (\ref{campoE1b}) and (\ref{campoE3b}) can be rewritten as
\begin{eqnarray}\label{campoE1c}
\textbf{E}^{(1)}(\textbf{x})|_{scatt}= \frac{1}{2\varepsilon_1}\int_{-\infty}^{+\infty}  \left[
\gamma^{(1)} H_1^{(1)}(k_{||}\rho) \hat{\rho}
+ik_{||}H_0^{(1)}(k_{||}\rho)\hat{z} 
 \right] A^{(1)}_p  \nonumber\\
\times e^{i \gamma^{(1)}z} k_{||}^2 dk_{||}, 
\end{eqnarray}
and 
\begin{eqnarray}\label{campoE3c}
\textbf{E}^{(3)}(\textbf{x})= \frac{1}{2 \varepsilon_1}\int_{-\infty}^{+\infty}  \left[
\gamma^{(1)} H_1^{(1)}(k_{||}\rho) \hat{\rho}
+i k_{||}H_0^{(1)}(k_{||}\rho)\hat{z} 
 \right] B^{(3)}_p \nonumber\\
\times e^{-i \gamma^{(1)}z} k_{||}^2 dk_{||}.
\end{eqnarray}
We now deform the integration path in (\ref{campoE1c}) and (\ref{campoE3c}) into a semicircle of large radius ($|k_{||}|\rightarrow \infty$) in the positive imaginary half--plane $\Im k_{||}>0$, avoiding the branch point and pole singularities, as indicated in Figure \ref{integracion}a. The vertical lines drawn from the branch point $k_1=\sqrt{\varepsilon_1}$ to $+i\infty$ ($-k_1$ to $-i\infty$) are the branch cut lines. 
%
%
Since, $H_n(k_{||} \rho) \approx e^{i k_{||} \rho}/\sqrt{k_{||}\rho}$ for $k_{||} \rho $  large enough, the contribution along that semicircle vanishes. The integration along the branch cut $B_1$ results in a   volume wave  \cite{wait,michalski}, which consist of a continuous spectrum of radiation modes. On the contrary, the residue contributions correspond to eigenmodes propagating in the radial direction and  with a discrete spectrum.  
In particular, we focus on distances between emitters smaller or of the same order than the propagation length of the waveguide eigenmodes. As a consequency,  the intensity of the electric field reached by the excitation of these modes are orders of magnitude larger than that corresponding to the excitation of the volume wave modes.
%
At these distances the energy transfer rate though eigenmodes is dominant and the volume wave mode contribution, which is in the order of the free space wave contribution, can be neglected. Then, the residues theorem gives
\begin{eqnarray}\label{campoE1d}
\textbf{E}^{(1)}(\textbf{x})|_{scatt}= \sum_j \textbf{E}^{(1)}_j(\textbf{x}) 
 , 
\end{eqnarray}
where
\begin{eqnarray}\label{campoE1dmodo}
\textbf{E}_j^{(1)}(\textbf{x})= \frac{1}{2\varepsilon_1} \left[
\gamma_j^{(1)} H_1^{(1)}(k_{||,j}\rho) \hat{\rho}
+i k_{||,j}H_0^{(1)}(k_{||,j}\rho)\hat{z} 
 \right] 
  e^{i \gamma_j^{(1)}z} k_{||,j}^2 \nonumber\\
 \times 2\pi i\, \mbox{Res} A^{(1)}_p.
\end{eqnarray}
Similarly, the application of the residues theorem in Eq (\ref{campoE3c}) gives

%
%
%
\begin{eqnarray}\label{campoE3d}
\textbf{E}^{(3)}(\textbf{x})=  \sum_j \textbf{E}^{(3)}_j
 , 
\end{eqnarray}
\begin{eqnarray}\label{campoE3dmodo}
\textbf{E}_j^{(3)}(\textbf{x})= \frac{1}{2 \varepsilon_1} \left[
\gamma_j^{(1)} H_1^{(1)}(k_{||,j}\rho) \hat{\rho}
+i k_{||,j}H_0^{(1)}(k_{||,j}\rho)\hat{z} 
 \right]  
 e^{-i \gamma_j^{(1)}z} k_{||,j}^2 \nonumber\\
\times 2\pi i \, \mbox{Res} B^{(3)}_p   
 , 
\end{eqnarray}
where $k_{||,j}$ is the propagation constant of a particular eigenmode, $j=$sp for SPs (ASPs or SSPs) or $j=$wg for WG modes (both quantities higher than the modulus of the photon wave vector in media 1 and 3), and Res is the
residue of the integrand in (\ref{campoE1c}) and (\ref{campoE3c}) at the pole $k_{||,j}$. 
\begin{eqnarray}\label{Res}
\mbox{Res} A_p^{(1)}= \lim_{k_{||}\to k_{||,j}}(k_{||}-k_{||,j}) A_p^{(1)}, \nonumber \\
\mbox{Res} B_p^{(3)}= \lim_{k_{||}\to k_{||,j}}(k_{||}-k_{||,j}) B_p^{(3)}. \nonumber \\
\end{eqnarray}
Inserting the expressions for $\textbf{E}^{(1)}(\textbf{x})|_{scatt}$ and $\textbf{E}^{(3)}(\textbf{x})$  given by Eqs. (\ref{campoE1d}) and (\ref{campoE3d}) into Eq. (\ref{ETn}) we obtain the following expression for the normalized energy transfer rate,
\begin{eqnarray}\label{ETnmodos}
F_{ET}=\sum_{lj} F_{lj} = \sum_{lj} \frac{T_{lj}(\omega)}{T_0(\omega)},
\end{eqnarray}
where 
\begin{eqnarray}\label{ETnmodo}
F_{lj}=\frac{T_{lj}(\omega)}{T_0(\omega)}=\frac{\hat{n}_A\cdot \textbf{E}_l(\textbf{x}_A) \, \hat{n}_A\cdot \textbf{E}_j^*(\textbf{x}_A) }{|\hat{n}_A\cdot \textbf{E}_0(\textbf{x}_A)|^2}.
\end{eqnarray}
Note that $F_{jj}$ is the contribution of the $j$ eigenmode channel to the normalized energy transfer rate  whereas $F_{l\not=j}$ corresponds to interference terms between $l$ and $j$ channels.

\subsubsection{Direct integration. Numerical quadrature}\label{integracion directa}

\begin{figure}
\centering
\resizebox{0.45\textwidth}{!}
{\includegraphics{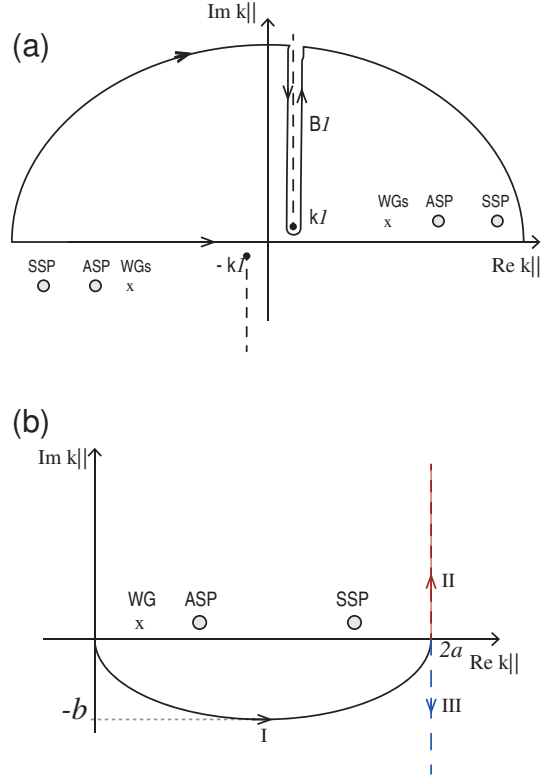}}
\caption{\label{fig:epsart} Singularities and path of integration in the complex plane $k_{||}=\Re k_{||}+i\Im{k_{||}}$ for the electromagnetic fields. (a) Integration path,  vertical branch cuts and poles captured. (b) The original path, along the positive semi--axis, is deformed along an elliptical path (I) surrounding the singularities, together with the paths (II) and (III) parallel to the imaginary $k_{||}$ axis using Hankel functions. 
}\label{integracion}
\end{figure}

In this subsection we transform the original oscillatory integrand function into one to avoid the complex singularities which lie near the real $k_{||}$ axis and then we apply a numerical quadrature to calculate the field integrals.  
This procedure does not  require to  determine the location of each pole, \textit{i.e.}, the determination of the complex propagation constant of the graphene waveguide. 
Unlike the method described in section \ref{modos grafeno} which 
allows to calculate separately the contribution of each one of the eigenmodes to the electric field, the present method only allows to calculate the total electric field scattered by the graphene waveguide.
To do this, we  follow a procedure similar to one developed in \cite{paulus}. We surround the pole  singularities by deforming the integration path into the complex plane as shown in Figure \ref{integracion}. The path I is an elliptical path starting at $k_{||}=0$ with the major semi--axis $k_{||}=a$ and the minor semi--axis $k_{||}=b$. In the region between  path I and the real axis the integrand function is analytical, thus the Cauchy's integral theorem implies that an integration on path I will be equal to the integral on the real axis from 0 to $2a$. The $a$ value should be chosen  large enough to surround  all pole singularities, thus $2a>\alpha_{\mbox{\tiny{ssp}}}$ must be fulfilled. Unlike the dielectric or  metallic waveguides, 
the propagation constant of the symmetric surface plasmon $\alpha_{\mbox{\tiny{ssp}}}$ on a graphene waveguide  
can reach values up to two orders of magnitude greater than that corresponding to a photon  of the same frequency.
Therefore, in a first step it is necessary  to divide the integration interval  into several subintervals and then  to apply the numerical quadrature in each subinterval. 

Taking into account the symmetry properties (\ref{bessel}) 
and the fact that the Hankel funtions of the first kind  $H_n^{(1)}(z)$ and the second kind $H_n^{(2)}(z)$  decrease faster as long as $|\Im z|$ increases in the sector $\Im z>0$ and $\Im z<0$, respectively, the remaining integration is carried out by deflecting the integration path from the real axis to a path parallel to the imaginary $k_{||}$ axis as shown in Figure 1,  with $\Im k_{||}>0$ for $H_n^{(1)}(k_{||}\rho)$ (path II) and with  $\Im k_{||}<0$ for $H_n^{(2)}(k_{||}\rho)$ (path III). In the region between path II and the real axis the integrand has no pole singularities, thus Cauchy's integral theorem implies that the integral on a closed path in this region   will be zero.Therefore, the integral over path II in the direction shown in Figure 1  is equal  to that from $2a$ to$+\infty$ over the real axis. In a similar way, one can demonstrate that the integral over path III in the direction shown in Figure 1  is equal  to that from $2a$ to$+\infty$ over the real axis. In our implementation, we have used a   32 point Gauss Legendre quadrature  to calculate the field integrals on paths I, II and III.  

\subsection{Case $\textbf{p}_D=p\hat{\rho}$} 
We next consider the case in which the donor dipole moment is oriented parallely to the graphene waveguide. Without loss of generality, we suppose this orientation in the $x$ axis. From Eq. (\ref{pym}), results,
\begin{eqnarray}\label{dparalelo}
d_p^{-}=\frac{\alpha \gamma^{(1)}}{\alpha^2+\beta^2}p \nonumber\\
d_s^{-}=-\frac{k_0 \gamma^{(1)}}{\alpha^2+\beta^2}p.
\end{eqnarray}
Following a similar way to that used to obtain Eqs. (\ref{campoE1}) and (\ref{campoE3}), we obtain the follow expressions for the electric field components below to the waveguide
\begin{eqnarray}\label{campoE3cx}
E^{(3)}(\textbf{x})|_{\rho}= \frac{i}{2\varepsilon_1}\int_{-\infty}^{+\infty}  \left[
- H_0^{(1)}(k_{||}\rho) + \frac{H_1^{(1)}(k_{||}\rho)}{\rho}
 \right] \cos \phi \, B^{(3)}_p \nonumber\\
\times e^{-i \gamma^{(1)}z} (\gamma^{(1)})^2 \, k_{||}  dk_{||}, 
\end{eqnarray}
\begin{eqnarray}\label{campoE3cy}
E^{(3)}(\textbf{x})|_{\phi}= \frac{i}{2\varepsilon_1}\int_{-\infty}^{+\infty}   [ (\gamma^{(1)})^2 \frac{H_1^{(1)}(k_{||}\rho)}{\rho} \, \sin \phi \, B^{(3)}_p  \nonumber \\
+\frac{k_1^2}{i} H_1^{(1)}(k_{||}\rho) \,  B^{(3)}_s]  \, e^{-i \gamma^{(1)}z} \,  k_{||} dk_{||}, 
\end{eqnarray}
\begin{eqnarray}\label{campoE3cz}
E^{(3)}(\textbf{x})|_{z}= -\frac{1}{2\varepsilon_1}\int_{-\infty}^{+\infty}  H_1^{(1)}(k_{||}\rho) \, B^{(3)}_p  \cos \phi  e^{-i \gamma^{(1)}z} \gamma^{(1)} k_{||}^2  dk_{||}, 
\end{eqnarray}
with similar expressions for the scattered electric field on the region located above the waveguide (medium 1). The residues theorem applied to these field components gives
\begin{eqnarray}\label{campoE3dx}
\textbf{E}^{(3)}(\textbf{x})=  \sum_j \textbf{E}^{(3)}_j
 , 
\end{eqnarray}
\begin{equation}\label{campoE3dmodox}
\begin{array}{ll}
\textbf{E}_j^{(3)}(\textbf{x})= \frac{i}{2 \varepsilon_1} [ \gamma_j^{(1)} \left(- H_0^{(1)}(k_{||,j} \rho) + \frac{H_1^{(1)}(k_{||,j} \rho)}{\rho} \right) \cos \phi \,\hat{\rho} \nonumber\\
+ \gamma_j^{(1)}  \frac{H_1^{(1)}(k_{||,j} \rho)}{\rho} \, \sin \phi \, \hat{\phi} + i \, k_{||,j} H_1^{(1)}(k_{||,j} \rho) \cos \phi \, \hat{z} ] \\  
\times e^{-i \gamma_j^{(1)}z} k_{||,j} \gamma_j^{(1)} \, 2\pi i \, \mbox{Res} B^{(3)}_p    
 , 
\end{array}
\end{equation}
where $k_{||,j}$ is the propagation constant of SPs, $j=$ASP, SSP.  Note that, for this dipole orientation, there are $s$ and $p$ polarized decay channels involved in the first and in the second term in Eq. (\ref{campoE3cy}), respectively. However, only $p$ polarized SPs exist in the graphene waveguide at the considered  frequency range, and as a consequence only the first term contribute  to the SP field. On the other hand, as in the vertical dipole orientation, we have neglected the contribution of the volume wave  electric field in Eq. (\ref{campoE3dx}) to the energy transfer  rate.

\section{Results}\label{resultados}

\begin{figure}[htbp!]
\centering
\resizebox{0.45\textwidth}{!}
{\includegraphics{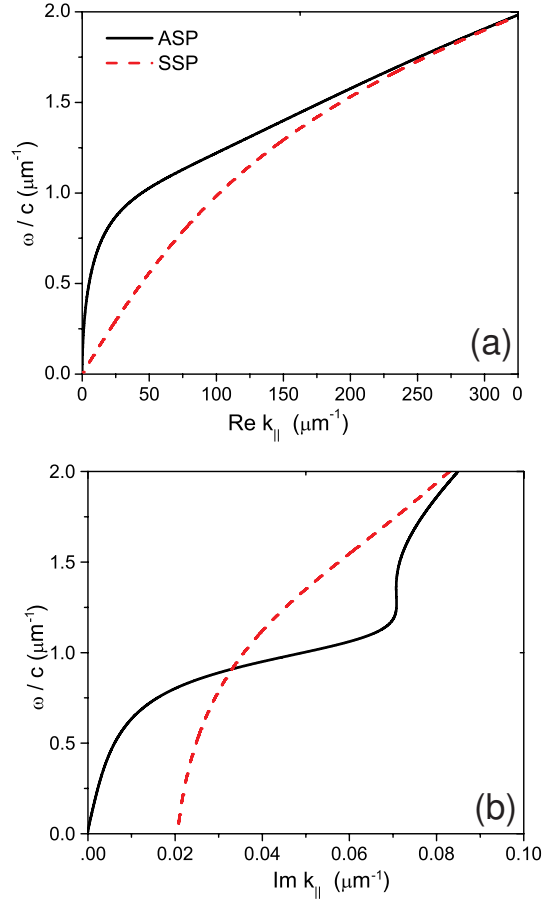}}
\caption{\label{fig:epsart} Dispersion curves for SSP and ASP modes, calculated for $\mu_c = 0.4$eV, $T=300$ K, $\gamma_c = 0.1$meV, $\varepsilon_1 = 1$ and $\varepsilon_2 = 3.9$. (a) $\Re k_{||}$ and (b) $\Im k_{||}$ as a function of $\omega/c$. 
}\label{disp}
\end{figure}

In this section we apply the formalism developed in  previous sections to calculate the energy transfer rate between two emitters localized close to a graphene waveguide. In order to obtain separate contributions of different eigenmodes, firstly we obtain the propagation constant of these modes by  requiring the denominator in the amplitudes $A_p^{(1)}$ and $B_p^{(3)}$ to be zero \cite{cuevas0}. 
Due to the high spatial confinement of SPs in comparison with that of the WG modes, the energy transfer through the graphene waveguide due to excitation of SPs is much greater than the corresponding energy transfer through the excitation of WG modes \cite{arruda2,cuevas0}. Thus, as we have verified,  the  ASP and SSP contributions dominate the energy transfer process on the frequency region presented in Figure \ref{disp} (where the SPs are well defined) and consequently the WG contributions can be neglected in Eq. (\ref{ETnmodos}). Since the imaginary part of the graphene conductivity (see \ref{grafeno}) $\Im \sigma$ changes sign from positive to negative at $\omega/\mu_c \approx  1.667$, due to the presence of the interband term in the conductivity,  $p$--polarized  SPs are well defined on the frequency range below such frequency. 

Figure \ref{disp} shows the propagation constant $k_{||}$ of the antisymmetric surface plasmon (ASPs) and symmetric surface plasmon (SSPs)  as a function of the frequency $\omega/c$ for $\mu_c=0.4$eV and $d=0.02\mu$m. With this value of $\mu_c$, $p$--polarized  SPs are supported by the structure for frequency values less than $\approx 3\mu$m$^{-1}$.  To appreciate the details of the dispersion SP curves, they are plotted in the range $\omega/c<2\mu$m$^{-1}$. 
Since the fields of these modes strongly overlap inside the thin layer (medium 2), the dispersion curves result in two well separated branches.  The upper branch corresponds to the ASP mode and the lower branch corresponds to the SSP mode. At high frequencies, $\omega/c>2\mu$m$^{-1}$,  SPs of the two graphene sheets are essentially uncoupled from each other and both the symmetric and the antisymmetric branches merge into the dispersion curve of the single SP mode supported by a graphene sheet.

Once the propagation constants are determined, the contribution of each SP modes to the total energy transfer rate has been calculated by the residues method (\ref{campoE1d}). 
Firstly, we briefly consider the case when the pair of donor and acceptor dipoles are on the same side of the graphene waveguide, and located at a distance $l$ from this one. Then, we consider the more realistic case when the donor and acceptor are placed on opposite sides on the waveguide. We focus on three configurations: the dipole moments are oriented perpendicular to the waveguide, the dipole moment of the donor is perpendicular to the waveguide and that of the acceptor is parallel to the waveguide, both dipole moments of donor and acceptor are parallel to the waveguide and parallel to each other, and both dipole moments of donor and acceptor are parallel to the waveguide and perpendicular to each other.
\begin{figure}[htbp!]
\centering
\resizebox{0.80\textwidth}{!}
{\includegraphics{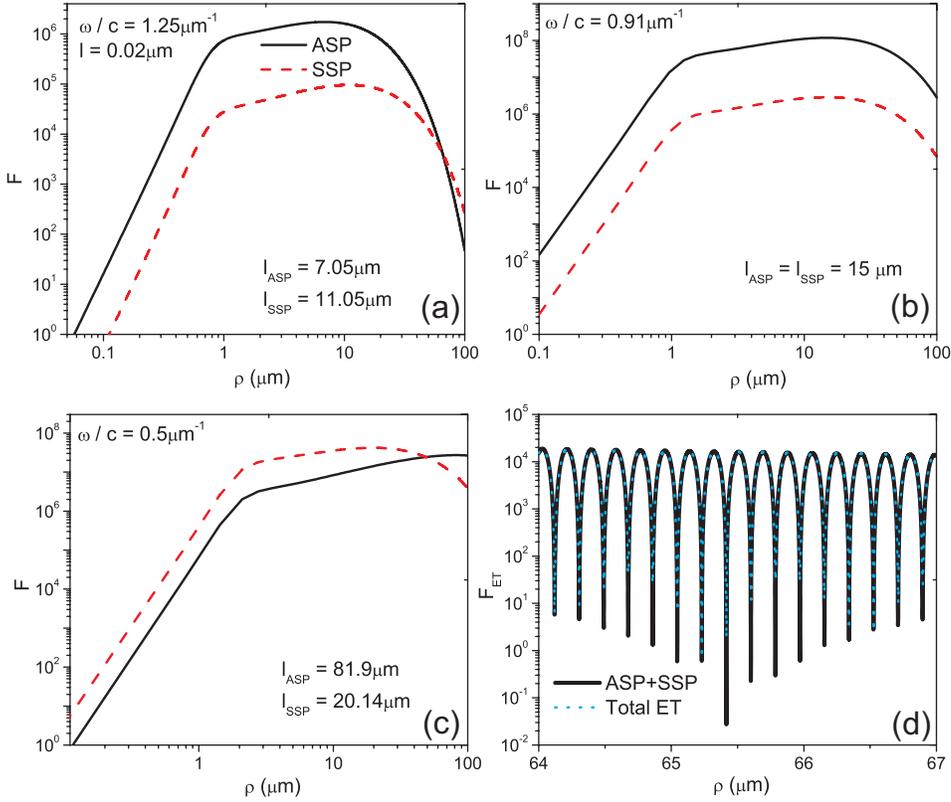}}
\caption{\label{fig:epsart} Contribution of the ASP and the SSP, $F_{\mbox{\tiny{ASP,ASP}}}$  and $F_{\mbox{\tiny{SSP,SSP}}}$, to the normalized energy transfer rate  between  two dipoles, both placed at the same $z=0.04\mu$m axis (at a distance $l=z_D-d=0.02\mu$m from the surface of the graphene waveguide), as a function of the in--plane separation $\rho$.  The frequency $\omega/c=1.25\mu$m$^{-1}$ (a), $\omega/c=0.91\mu$m$^{-1}$ (b) and $\omega/c=0.5\mu$m$^{-1}$ (c), and the dipole moments are oriented along the $z$ axis.  (d) Normalized total energy transfer rate and the superposition of the ASP and SSP as a function of the in--plane distance between the donor and acceptor.  The waveguide parameters are the same as in Figure \ref{disp}.
}\label{omegac1p25}
\end{figure}
%

In Figure \ref{omegac1p25} we have plotted the contribution of the SP modes to the normalized energy transfer rate, \textit{i.e.}, the energy transfer terms in (\ref{ETnmodo}) 
with $l=j=$ASP and $l=j=$SSP ($F_{\mbox{\tiny{ASP,ASP}}}$ and $F_{\mbox{\tiny{SSP,SSP}}}$), as a function of the in--plane separation between the donor and acceptor and for several $\omega/c=1.25, \, 0.91, \, \mbox{and} \, 0.5 \mu$m$^{-1}$  frequency transition values. 
The donor and acceptor are localized on the same side of the waveguide (medium 1), at $\textbf{x}_D=0.04\mu$m$\hat{z}$ and $\textbf{x}_A=\rho \hat{\rho}+0.04\mu$m$\hat{z}$, respectively, and with their dipole moments aligned along the $z$ axis. 
In all the cases, the normalized energy transfer contributions is larger than 10, pointing out that SP excitations dominate the energy transfer process in the distance range considered  in Figure \ref{omegac1p25}. At larger distances the imaginary part of the  SPs propagation constant turns out an exponentially decaying field and, as a consequence the energy transfer  begins to be dominated by the free--space radiation [not shown in Figure \ref{omegac1p25}]. As in the single graphene sheet case, we have verified that the  crossover distance occurs at $\approx 10 L-20 L$ \cite{nikitin} where $L=\frac{1}{2 \Im\,\alpha_{\mbox{\tiny{SP}}}}$ is the propagation length of SPs.

In Figure \ref{omegac1p25}a, we observe that the maximum value of the normalized energy transfer rate for the SSP falls at $\rho_{max} \approx 11\mu$m, a value that is larger than that corresponding to the ASP, $\rho_{max} \approx 7\mu$m.    
On the other hand, 
by using the calculated $\Im\,\alpha_{\mbox{\tiny{SP}}}$ values plotted 
 in Figure \ref{disp}b for $\omega/c=1.25\mu$m$^{-1}$, we have obtained $L_{\mbox{\tiny{SSP}}}=11.05\mu$m and $L_{\mbox{\tiny{ASP}}}=7.05\mu$m for the SSP and the ASP, respectively, which agree well with $\rho$ values where the energy transfer reaches its maximum value. The correspondence between the position of these maxima and the propagation length of SPs, which has been reported in \cite{biehs} for the case of single graphene sheet, can be understood 
as follows: since the $\hat{z}$--component of the SP electric field (\ref{campoE1dmodo}) depends on the $\rho$ distance as $E_{\mbox{\tiny{sp}},z}(\rho) \approx H_0^{(1)}(\alpha_{\mbox{\tiny{SP}}} \rho)$, for argument values large enough, $\alpha_{\mbox{\tiny{SP}}}\rho>>1$, it follows that 
\begin{equation} \label{Esp}
E_{\mbox{\tiny{SP}},z}(\rho)  \approx \frac{e^{i \alpha_{\mbox{\tiny{SP}}} \, \rho}}{\sqrt{\rho}}= \frac{e^{i \Re \alpha_{\mbox{\tiny{SP}}} \, \rho-\Im \alpha_{\mbox{\tiny{SP}}} \, \rho}}{\sqrt{\rho}}.
\end{equation}         
%
Taking into account that, in the absence the graphene  waveguide, the field of the donor is written as (refer to \ref{ap1} for its derivation)
\begin{equation}
E_{0,z}(\rho) = \frac{e^{i k_1 \rho}}{\rho}, 
\end{equation}         
it follows that the energy transfer contribution of either of the two SP channels (\ref{campoE1dmodo}) can be written as
\begin{equation}
F_{\mbox{\tiny{SP}}} = |\frac{E_{\mbox{\tiny{SP}},z}(\rho) }{E_{0,z}(\rho) }|^2 =\rho e^{-2 \Im \alpha_{\mbox{\tiny{SP}}}\,\rho}, 
\end{equation}         
which reaches its maximum value at $\rho=\frac{1}{2\Im \alpha_{\mbox{\tiny{SP}}}}$. 
From this fact and taking into account the values of $\Im \alpha_{\mbox{\tiny{SP}}}$ calculated in Figure \ref{disp}b, we conclude that for frequency values less (greater) than $0.91\mu$m$^{-1}$, the curve of  $F_{\mbox{\tiny{ASP,ASP}}}$ reaches its maximum value at a distance greater (less) than that  corresponding to the curve of $F_{\mbox{\tiny{SSP,SSP}}}$.
Figures \ref{omegac1p25}a--c confirm such behavior.

It is worth noting that the energy transfer rate between the donor and acceptor arises from a coherent  superposition of the   
ASP and the SSP contributions (both calculated in Figures \ref{omegac1p25}a--c) into Eq. (\ref{ETnmodos}), leading to a spatial modulation due to interference terms $F_{lj}$ with $l=\mbox{ASP},\,j=\mbox{SSP}$ and $l=\mbox{SSP},\,j=\mbox{ASP}$. This fact is illustrated in Figure \ref{omegac1p25}d where we have plotted the normalized energy transfer  rate for $\omega/c=1.25\mu$m$^{-1}$ 
by considering only the SP terms in Eq. (\ref{ETnmodos}),  
on the one hand,  
 and by direct integration, \textit{i.e.}, by using Eq. (\ref{campoE1c}) as explained in subsection \ref{integracion directa}, on the other hand. We see that both curves match, 
confirming that the energy transfer rate is well approximated by the coherent superposition of both ASP and SSP mode  channels  
provided that the in--plane separation between the donor and acceptor  is comparable or even lower than the SP propagation length. 
In order to  visualize a strong  interference  effect,  
in Figure \ref{omegac1p25}c  we have plotted the energy transfer curve for a distance range close to $65\mu$m  in which the ASP and the SPP   energy transfer contributions are approximately equals to each other, as can be seen in Figure \ref{omegac1p25}a. We observe a pronounced spatial oscillation 
whose  period $\Lambda \approx 0.185\mu$m, a value that can be calculated 
within the framework of the model in Eq. (\ref{ETnmodos}), where the interference term between the SSP and ASP is written as   
\begin{equation}\label{Flj}
I=F_{\mbox{\tiny{SSP}},\mbox{\tiny{ASP}}}+F_{\mbox{\tiny{ASP}},\mbox{\tiny{SSP}}} \approx \cos [(\Re \alpha_{\mbox{\tiny{SSP}}}-\Re \alpha_{\mbox{\tiny{ASP}}})\rho].
\end{equation}         
Equation (\ref{Flj}) shows a periodic spatial dependence along $\rho$ direction with a period  $\Lambda=2 \pi / (\Re \alpha_{\mbox{\tiny{SSP}}}-\Re \alpha_{\mbox{\tiny{ASP}}})=2 \pi/(142.2\mu\mbox{m}^{-1}-108.2\mu\mbox{m}^{-1}) \approx 0.1847\mu$m, where we have used the values of the  SP propagation constants calculated in Figure \ref{disp}a for  $\omega/c=1.25\mu$m$^{-1}$.

We now consider the SP coupling efficiency supporting the energy transfer between donor and acceptor on opposite sides of the graphene waveguide. As in the metallic slab  case \cite{barnes,pra84}, we expect that the electromagnetic field extending between the two graphene layers of the waveguide being able to facilitate the energy transfer between dipoles placed on opposite sides of the waveguide. 
\begin{figure}
\centering
\resizebox{0.45\textwidth}{!}
{\includegraphics{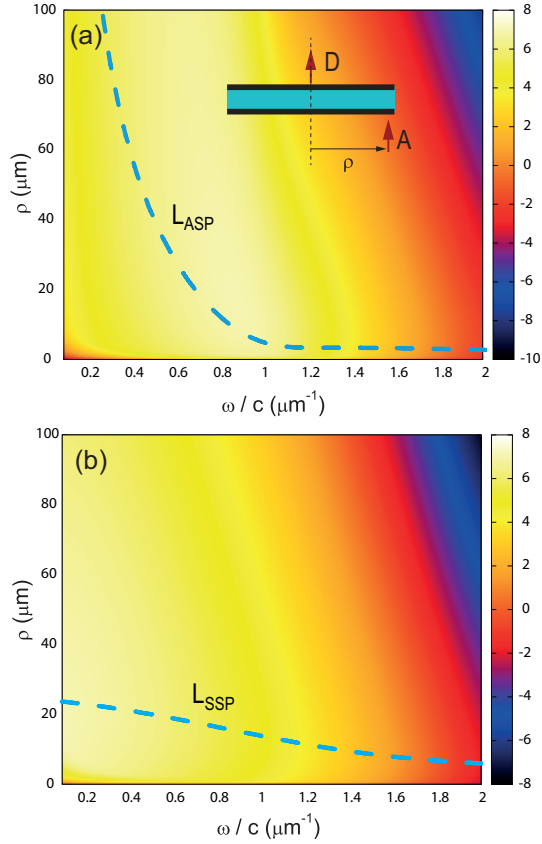}}
\caption{\label{fig:epsart} Map of the  SP contributions $F_{\mbox{\tiny{ASP,ASP}}}$ (a) and $F_{\mbox{\tiny{SSP,SSP}}}$ (b) as a function of frequency and the in--plane distance between the donor and acceptor. The dashed line curves correspond to the propagation lengths of SPs. The donor and acceptor are localized on opposite sides of  the waveguide, $z_D=0.04\mu$m (at a distance $l=0.02\mu$m from the top of the waveguide) and $z_A=-0.02\mu$m (at a distance $l=0.02\mu$m from the bottom of the waveguide), with their dipole moments   along the $z$ axis. The waveguide parameters are the same as in Figure \ref{disp}. 
}\label{mapa}
\end{figure}
Our calculations confirm this expectation, as can be seen in Figure \ref{mapa} where we show plots of the 
SP  contributions to the normalized energy transfer rate, $F_{\mbox{\tiny{ASP,ASP}}}$ and $F_{\mbox{\tiny{SSP,SSP}}}$,  as a function of the in--plane distance $\rho$  and   $\omega/c$ frequency for the case where the donor is placed  at $\textbf{x}_D=0.04 \mu $m $\hat{z}$ (above the waveguide) and the acceptor   is placed at $\textbf{x}_A=\rho \hat{\rho}-0.02\mu$m$\hat{z}$ (below the waveguide). The electric dipole of the donor and acceptor  are oriented along the $z$ axis. As in the previously presented case where the donor  and acceptor were localized on the same side of the waveguide, we also observe that  the maximum normalized energy transfer rate is obtained for a value of the in--plane separation close to the  propagation length of the SP. 

\begin{figure}[htbp!]
\centering
\resizebox{0.45\textwidth}{!}
{\includegraphics{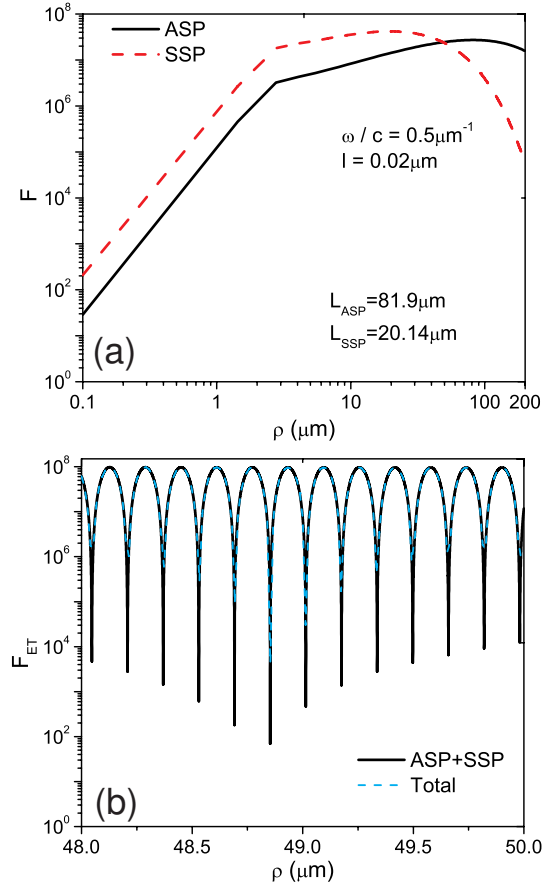}}
\caption{\label{fig:epsart} Contribution of the ASP and the SSP, $F_{\mbox{\tiny{ASP,ASP}}}$  and $F_{\mbox{\tiny{SSP,SSP}}}$ (a), to the normalized energy transfer rate  between D and A placed on opposite sides of the waveguide as a function of the $\rho$ distance between them.  (b) Normalized total energy transfer rate and the superposition of the ASP and SSP as a function of the distance between D and A dipoles.  The transition  frequency $\omega/c=0.5\mu$m$^{-1}$ and  the dipole moments are oriented along the $z$ axis. The waveguide parameters and the location of D and A are the same as in Figure \ref{mapa}. 
}\label{omegac0p5}
\end{figure}

To appreciate the spatial details of the SP energy transfer from donor to acceptor, in Figure \ref{omegac0p5}a we have plotted the ASP and SSP contributions as a function of the in--plane  $\rho$ separation for $\omega/c=0.5\mu$m$^{-1}$.  
Since the frequency value $\omega/c<1\mu$m$^{-1}$,  the ASP contribution curve exhibits its maximum value at an in--plane separation greater than that corresponding to the SSP contribution curve ($\rho_{max,\mbox{\tiny{ASP}}}=81.9\mu$m and $\rho_{max,\mbox{\tiny{SSP}}}=20.14\mu$m). 
Figure \ref{omegac0p5}b shows the  normalized total energy transfer rate $F_{ET}$ calculated by direct integration as it  has been described in section \ref{integracion directa}, and by the coherent superposition of the ASP and SSP channels to the normalized energy transfer (\ref{ETnmodos}), for distance values close to $49\mu$m (where the ASP and SSP energy transfer contributions are approximately the same). We observe a great 
spatial modulation due to the interference effect between the ASP and SSP whose period $\Lambda \approx 0.16\mu$m, a value that can be obtained by 
 using the  propagation constant values shown in Figure 3a  for $\omega/c=0.5\mu$m$^{-1}$, 
$\Lambda=2 \pi / (\Re \alpha_{\mbox{\tiny{SSP}}}-\Re \alpha_{\mbox{\tiny{ASP}}})=2 \pi/(44.055\mu$m$^{-1}-5.085\mu$m$^{-1}) \approx 0.1612\mu$m$^{-1}$.

\begin{figure}[htbp!]
\centering
\resizebox{0.45\textwidth}{!}
{\includegraphics{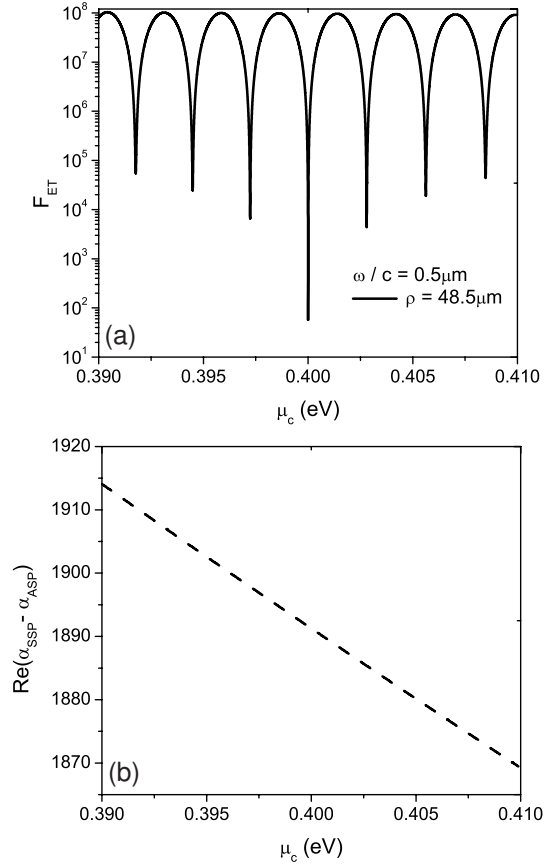}}
\caption{\label{fig:epsart} (a) Total energy transfer rate $F_{ET}$ between donor and acceptor placed on opposite sides of the waveguide as a function of the chemical potential $\mu_c$. (b) $\Delta \alpha_{\mbox{\tiny{SP}}}(\mu_c)$  for $\mu_c$ values near to $\mu_c=0.4$eV.  
The frequency $\omega/c=0.5\mu$m$^{-1}$,  the dipole moments are oriented along the $z$ axis and the in--plane separation is set at $\rho=48.5\mu$m. The waveguide parameters are the same as in Figure \ref{omegac0p5}. 
}\label{amu}
\end{figure}

To further investigate the effect that the variation of the chemical potential has on the energy transfer rate, we specify the in--plane separation between the donor and acceptor and vary the chemical potential $\mu_c$. Even though $\mu_c$ variations are manifested as a significant energy transfer modification   in the whole range of the donor--acceptor separation considered in Figure \ref{omegac0p5}, for the sake of clarity we have chosen $\rho=48.85\mu$m, a value 
for which the total energy transfer rate exhibits an absolute minimum value (Figure \ref{omegac0p5}b).    
Figure \ref{amu} shows a plot of the normalized total energy transfer $F_{ET}$ as a function of the chemical potential  $\mu_c$ for the same configuration as in Figure \ref{omegac0p5}. 
We observe that, by varying the chemical potential from $0.4$eV, the normalized energy transfer rate  notably increases, near six orders of magnitude,  towards values close to $10^8$. Then, $F_{ET}$ decreases towards another minimum and follows  the same behavior, almost periodically, as the chemical potential modulus is increased. The period of the oscillations,  determined by the distance between two consecutive minima in  Figure \ref{amu}a, is $\Lambda_\mu \approx 0.0028$eV. The origin of this periodic behavior is explained by Eq. (\ref{Flj}) where the argument of the interference term $\varphi(\mu_c)=\Delta \alpha_{\mbox{\tiny{SP}}}(\mu_c)\,\rho$ ($\Delta \alpha_{\mbox{\tiny{SP}}}= \Re \alpha_{\mbox{\tiny{SSP}}}(\mu_c)-\Re \alpha_{\mbox{\tiny{ASP}}}(\mu_c)$) is a function of $\mu_c$. In Figure \ref{amu}b we have plotted $\Delta \alpha_{\mbox{\tiny{SP}}}$ as a function of $\mu_c$ in the range $0.39$eV$<\mu_c<0.41$eV. In this range,  $\Delta \alpha_{\mbox{\tiny{SP}}}$ is a linear function of $\mu_c$ and as a consequence the argument can be approximated by $\varphi(\mu_c)=\varphi(\mu^{(0)}_c)+\Delta \alpha_{\mbox{\tiny{SP}}}'(\mu^{(0)}_c)\,(\mu_c-\mu^{(0)}_c) \, \rho$, where $\mu^{(0)}_c=0.4$eV, $\Delta \alpha_{\mbox{\tiny{SP}}}'(\mu^{(0)}_c)=\Delta \alpha_{\mbox{\tiny{SP}}}'(0.4$eV$)$ 
is the derivative with respect to $\mu_c$ and $\varphi(\mu^{(0)}_c)=\Delta \alpha_{\mbox{\tiny{SP}}}(0.4$eV$)\,\rho$. Therefore, the period of the interference function (\ref{Flj}) is $\Lambda_\mu=2\pi/[\Delta \alpha_{\mbox{\tiny{SP}}}'(0.4$eV$)\,\rho]$. 
From Figure \ref{amu}b we determine $\Delta \alpha_{\mbox{\tiny{SP}}}'(0.4$eV$)=-46.125\mu$m$^{-1}$eV$^{-1}$, thus 
for $\rho=48.853\mu$m 
it follows that $\Lambda_\mu=2\pi/2253.23\,$eV$ \approx 0.002788 $eV, a value that agrees well with that previously determined  value.   

\begin{figure}[htbp!]
\centering
\resizebox{0.45\textwidth}{!}
{\includegraphics{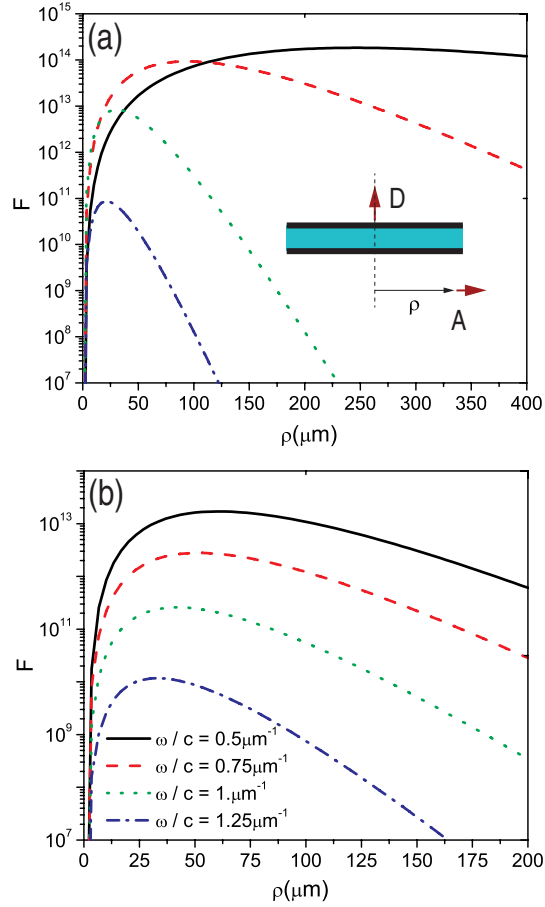}}
\caption{\label{fig:epsart} ASP and SSP contributions, $F_{\mbox{\tiny{ASP,ASP}}}$ (a)  and $F_{\mbox{\tiny{SSP,SSP}}}$ (b), to the normalized energy transfer rate  between D and A placed on opposite sides of the waveguide as a function of the $\rho$ distance between them. The dipole moments are oriented along the $z$ axis (donor) and  along the $x$ axis (acceptor). The waveguide parameters and the location of D and A are the same as in Figure \ref{mapa}. 
}\label{dzdx}
\end{figure}

We next consider the configuration in which the  orientation of the acceptor dipole is along the $\rho$ axis as it is shown in the inset of Figure \ref{dzdx}a. 
We observe that both $F_{\mbox{\tiny{ASP,ASP}}}$ and $F_{\mbox{\tiny{SSP,SSP}}}$ curves exhibit a maximum value at approximately the same in--plane separation $\rho$ when frequency is $\omega/c=1\mu$m$^{-1}$. 
However, unlike the previous configuration (both donor and acceptor 
dipole moments aligned along the $z$ axis),   
the distance $\rho_{max}(1\mu$m$^{-1}) \approx 35 \mu$m, a value representing threefold the SP 
propagation length ($L(1\mu$m$^{-1}) \approx 12\mu$m).  
In addition, we have verified the same behavior for all frequencies. This fact can be understood by taking into account that, in the absence of the graphene waveguide, the $\hat{x}$--component of the donor electric field is written as (see \ref{ap1})
\begin{equation}
E_{0,x}(\rho) = \frac{e^{i k_1 \rho}}{\rho^2}, 
\end{equation} 
while the $\hat{x}$--component of the SP electric field is as in Eq. (\ref{Esp}), $E_{\mbox{\tiny{SP}},x}(\rho) \approx E_{\mbox{\tiny{SP}},z}$. Thus, the energy transfer contribution of the SP channel (\ref{campoE3dmodo}) can be written as
\begin{equation}
F_{\mbox{\tiny{SP}}} = |\frac{E_{\mbox{\tiny{SP}},x}(\rho) }{E_{0,x}(\rho) }|^2 =\rho^3 e^{-2 \Im \alpha_{\mbox{\tiny{SP}}}\,\rho}, 
\end{equation}         
which reaches its maximum value at $\rho=\frac{3}{2\Im \alpha_{\mbox{\tiny{SP}}}}=3 L_{\mbox{\tiny{SP}}}$.

\begin{figure}
\centering
\resizebox{0.80\textwidth}{!}
{\includegraphics{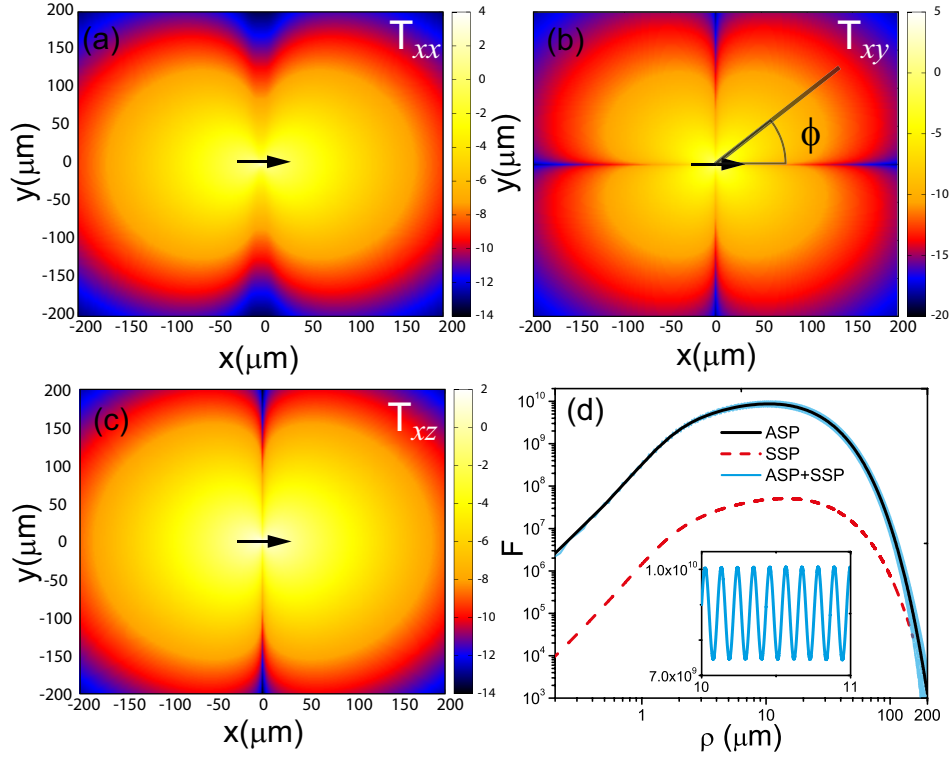}}
\caption{\label{fig:epsart} ASP contribution, $F_{\mbox{\tiny{ASP,ASP}}}$, to the  energy transfer function $T$ (in logarithmic scale)  between D and A placed on opposite sides of  the waveguide ($\textbf{x}_D=0.04\mu$m$\hat{z}$ and $\textbf{x}_A=x \hat{x}+y\hat{y}-0.02\mu$m$\hat{z}$)  as a function of the $x$ and $y$ axis on the $z=-0.02\mu$m plane (where A is placed). The dipole moment of D is along the $x$ axis  and the dipole moment of A is along the $x$ axis (a), the $y$ axis (b) and the $z$ axis (c). (d) ASP and SSP contributions to the normalized energy transfer rate  as a function of the distance $\rho$ along the line at $\phi=45^\circ$ in Figure \ref{mapax}b. 
The waveguide parameters and the location of D and A are the same as in Figure \ref{mapa}. 
}\label{mapax}
\end{figure}

In addition to considering the energy transfer rate for the donor dipole moment aligned along the $z$ axis, 
another useful configuration is that in which the donor electric dipole lies parallel to the graphene waveguide. 
The acceptor is placed at the $z=-0.02\mu$m$^{-1}$ plane (below the waveguide). The position of the donor is fixed at $\textbf{x}_D=0.04\mu$m$\hat{z}$ (above the waveguide) and its projection on the acceptor plane is indicated with an arrow in Figure \ref{mapax}. Without loss of generality we have chosen the donor orientation along the $x$ axis ($\textbf{p}_D=p \hat{x}$).  
Figure \ref{mapax} shows the spatial distribution of the ASP contribution to the energy transfer function $T$  as a function of the acceptor position $(x,\,y)$ on $z=-0.02\mu$m$^{-1}$ plane. The frequency transition of the donor is chosen to be $\omega/c=1\mu$m$^{-1}$. 
%
The color map is calculated for the acceptor placed in each point of the plot. 
Unlike the case that we previously studied (when the donor orientation is parallel to the $z$ axis) for which both the transfer function $T$ and the unbounded  transfer function $T_0$ are invariant under rotation around the $z$ axis, in the present case neither $T$ nor $T_0$ are invariant under rotation around the $z$ axis, although the $T/T_0$ rate (and as a consequence also $F$) is itself invariant. For this reason, and to highlight the symmetry of the energy transfer in the acceptor  plane, 
we believe that the calculation of $T$ is particularly appropriate. 

From
Figure \ref{mapax}  
we can see
that  
the energy transfer rate presents a strongly angular
dependence. For instance, Figure \ref{mapax}b  shows four zones on the $z=-0.02\mu$m plane in which the energy transfer rate is enhanced. These zones appear limited by two lines $x=0$ and $y=0$ where the component of the donor electric field that lies parallel to the acceptor plane  is in the $\hat{x}$ direction. As a consequence the energy transfer to the acceptor positioned at these lines with its dipole moment along the $\hat{y}$ direction is zero as can be seen in Figure \ref{mapax}b. 
On the other hand, the energy transfer to the acceptor oriented in the $z$ axis  presents two enhanced zones limited by the $x=0$ line (Figure \ref{mapax}c), where the $\hat{z}$ component of the  donor electric field is zero. 

Since the imaginary part of the propagation constant value of the SSP at $\omega/c=1\mu$m$^{-1}$ is approximately the same as the ASP,  the map corresponding to the SSP [not shown in Figure \ref{mapax}]  is similar to that in Figure \ref{mapax} for the ASP. 

In order to highlight the role of SPs in the energy transfer rate, we calculate the ASP and the SSP contributions to the normalized energy transfer $F_{ET}$   along the line in Figure \ref{mapax}b with  $\phi=45^\circ$. The acceptor dipole moment lies parallel to the $y$ axis. Figure \ref{mapax}d  shows that the maximum value of the ASP contribution curve is near  three orders of magnitude larger than the value obtained for the  SSP contribution, a fact occurring in the whole acceptor plane regardless the $\phi$ angle. 
Both ASP and SSP curves exhibit their maximum values almost at the same $\rho_{max}$ distance, a fact that follows from the similarity between the imaginary parts of the  ASP and SSP propagation constant at $\omega/c = 1\mu$m$^{-1}$. Furthermore, this value is $\rho_{max}\approx 12\mu$m, which agrees well with that of the SP propagation length at this frequency. This is true because the $\hat{y}$ component of the SP electric field (ASP or SSP) given by Eq. (\ref{campoE3dmodox}) for $\phi=45^\circ$ and for long argument is $E_{\mbox{\tiny{SP}},y}(\rho) \approx \frac{e^{i \alpha_{\mbox{\tiny{SP}}} \, \rho}}{\sqrt{\rho}}$, whereas the $\hat{y}$ component of the electric field for the same dipole in an unbounded medium is $E_{0,y}(\rho) \approx \frac{e^{i k_1 \rho}}{\rho}$. Therefore, $F = \frac{|E_{\mbox{\tiny{SP}},y}|^2}{|E_{0,y}(\rho)|^2} \approx \rho e^{-2 \Im \alpha_{\mbox{\tiny{SP}}}\,\rho}$ %
which reaches its maximum value at $\rho=\frac{1}{2\Im \alpha_{\mbox{\tiny{SP}}}}= L_{\mbox{\tiny{SP}}}$. We have verified [not shown in Figure \ref{mapax}] that in the case that the electric dipole of the  acceptor is set along the $z$ axis both ASP and SSP contribution curves exhibit their  maximun values at  $\rho_{max}\approx 35\mu$m, a value threefold larger than the SP propagation length $L(1\mu$m$)=12\mu$m. This fact can be explained by taking into account that the $\hat{z}$ component of the SP electric field (ASP or SSP) given by Eq. (\ref{campoE3dmodox}) for $\phi=45^\circ$ and for long argument is $E_{\mbox{\tiny{SP}},z}(\rho) \approx \frac{e^{i \alpha_{\mbox{\tiny{SP}}} \, \rho}}{\sqrt{\rho}}$, whereas the $\hat{z}$ component of the electric field for the same dipole in an unbounded medium is $E_{0,z}(\rho) \approx \frac{e^{i k_1 \rho}}{\rho^2}$.   Thus, $F = \frac{|E_{\mbox{\tiny{SP}},z}|^2}{|E_{0,z}(\rho)|^2} \approx \rho^3 e^{-2 \Im \alpha_{\mbox{\tiny{SP}}}\,\rho}$ which reaches its maximum value at $\rho=\frac{3}{2\Im \alpha_{\mbox{\tiny{SP}}}}= 3 L_{\mbox{\tiny{SP}}}$. 
 We also observe interference oscillations in 
the total energy transfer curve, which agree well with the curve  obtained by the coherent superposition between ASP and SSP modes.  At low distances, where the main contribution to the energy transfer rate comes from  the ASP, the interference effects are less pronounced than the corresponding effects at large distances, where the ASP and SSP contributions are similar. To appreciate the details of the  interference between SP channels, in  the inset of Figure \ref{mapax}d we have enlarged the horizontal scale of such figure. In the inset it is clearly shown that the oscillations present a spatial period $\Lambda \approx 0.11\mu$m, a value that agrees well with that calculated by using the real parts of SPs obtained in Figure \ref{disp}a for $\omega/c=1\mu$m$^{-1}$, $\Lambda=2\pi/(\Re \alpha_{\mbox{\tiny{SSP}}}-\Re \alpha_{\mbox{\tiny{SSP}}}) \approx 2\pi/(102.4\mu \mbox{m}^{-1}-44.4\mu \mbox{m}^{-1})=0.108\mu$m.

\section{Conclusions} \label{conclusiones}

The energy transfer rate from a donor to an acceptor close to a planar graphene waveguide has been studied by
 applying an analytical classical method enabling to obtain a rigorous solution in a closed integral form. We have developed two methods to perform the field integration. With the first method (method \textit{i}) we have  extracted  SP  contributions to the electromagnetic field of the integral solutions, whereas with the  the second method (method \textit{ii}) we have numerically solved the field integrals by deforming the path of integration into a suitable path in the complex plane.

We have considered the donor and acceptor pair on the same side of the graphene waveguide as well as the case when  the donor and acceptor are placed on opposite sides on the waveguide. By using the method \textit{i} we separately calculated the contribution of symmetric and antisymmetric SPs to the total energy transfer rate. On the other hand,  we have calculated the normalized energy transfer rate by direct integration (by using the method \textit{ii}) and we have verified that the total energy transfer rate is  successfully approximated by a coherent superposition of the two symmetric and antisymmetric mode channels for distances 
of the  same order as that of the SP propagation lengths.

In the presented examples, we have varied the distance between donor and acceptor as well as their dipole moment orientations. 
Our calculations show that the symmetric and antisymmetric maximum contributions to the total normalized energy transfer are reached as the donor--acceptor distance approaches one of  the SP propagation lengths for the configuration when both dipole moments are oriented perpendicular or parallel to the waveguide. On the other hand, for the configuration where the dipole moment of the donor (acceptor) is perpendicular to the waveguide and that of the acceptor (donor)  
is parallel to the waveguide  these maximum contributions are reached at a distance three times larger than the SP propagation lengths. 

The possibility of varying the chemical potential of the graphene sheets constitutes a degree of freedom   allowing to modify symmetric or antisymmetric SP branches, and their influence on the energy transfer rate. Our calculations have shown strong  spatial variations with the chemical potential, with a modulation period being inversely proportional to the derivative of the difference between the symmetric and antisymmetric SP propagation constants. It has been an objective of the present paper the determination of the SP propagation constant, which is usually considered as a tedious task, to provide  a comprehensive analysis about the role that such modes play in the modification of the normalized energy transfer rate from donor to acceptor. 


\section*{Acknowledgment}
The authors acknowledge the financial support of Consejo Nacional de Investigaciones Cient\'{\i}ficas y T\'ecnicas (CONICET). 


\appendix

\section{Electric field of a dipole in an unbounding medium} 
\label{ap1}
\setcounter{equation}{0}
\renewcommand{\theequation}{A{\arabic{equation}}}

We consider an electric dipole at position $\textbf{x}_D$ oriented along $\hat{z}$ direction, $\textbf{p}=p\hat{z}$. The vector potential $\textbf{A}_e(\textbf{x})$ in spherical coordinates is given by \cite{novotny}
\begin{equation}\label{A_e_esfericas}
\textbf{A}_e(\textbf{x})=\frac{ k_0\,e^{i k_1 R}}{i R} p \hat{z},
\end{equation}  
where $R=|\textbf{x}-\textbf{x}_D|$. By using Eq. (\ref{camposEHp}), we obtain the follow components for the electric field in cartesian coordinates

\begin{eqnarray}
\begin{array}{cc}
\label{Elibre}
E_{x}(\textbf{x})=
\frac{ e^{i k_1 R}}{\varepsilon_1 R} \left[\left(i k_1-\frac{1}{R}\right)^2+\left(-\frac{i k_1}{R}+\frac{2}{R^2}\right)\right]\frac{z-z_D}{R} \frac{x-x_D}{R},\\
E_{y}(\textbf{x})=
\frac{ e^{i k_1 R}}{\varepsilon_1 R} \left[\left(i k_1-\frac{1}{R}\right)^2+\left(-\frac{i k_1}{R}+\frac{2}{R^2}\right)\right]\frac{z-z_D}{R} \frac{y-y_D}{R},\\
E_{z}(\textbf{x})=
\frac{ e^{i k_1 R}}{\varepsilon_1 R} \\
\times \left\{\left[\left(i k_1-\frac{1}{R}\right)^2+\left(-\frac{i k_1}{R}+\frac{2}{R^2}\right)\right]\frac{(z-z_D)^2}{R^2}+\left(i k_1-\frac{1}{R}\right)\frac{1}{R}\right\}\\
+k_0^2 \frac{ e^{i k_1 R}}{R}.
\end{array}
\end{eqnarray}  
For the electric dipole orientation along the  $\hat{x}$ direction, $\textbf{p}=p\hat{x}$, the field components can be obtained from Eqs. (\ref{Elibre}) by replacing $E_x \rightarrow E_z$, $E_z \rightarrow E_x$, $x \rightarrow -z$ and $z \rightarrow x$.  

\section{Electric field of a dipole  in presence of a graphene waveguide} 
\label{ap2}
\setcounter{equation}{0}
\renewcommand{\theequation}{B{\arabic{equation}}}

Taking into account the infinitesimal translational invariance in the $x$ and $y$ directions, 
the  field of the electric dipole $p_D$ can be represented as a superposition of two  basic polarization modes: $p$ polarization mode, for which the magnetic field  is parallel to the $x-y$ plane in Figure \ref{sistema}, and $s$ polarization mode, for which the electric field is parallel to the  $x-y$ plane. From the mathematical point of view,  the electromagnetic field can be represented by two scalar functions $a_p(\textbf{x})$ and $a_s(\textbf{x})$ which are, respectively, the $\hat{z}$ component electric and magnetic vector potentials \cite{cuevas0,novotny},
\begin{eqnarray}
\textbf{A}_e(\textbf{x})=a_p(\textbf{x})\,\hat{z}, \nonumber \\
\textbf{A}_h(\textbf{x})=a_s(\textbf{x})\,\hat{z}. \label{ApAs}
\end{eqnarray}
The electric field $\textbf{E}=\textbf{E}_p(\textbf{x})+\textbf{E}_s(\textbf{x})$ and the magnetic field $\textbf{H}=\textbf{H}_p(\textbf{x})+\textbf{H}_s(\textbf{x})$ can be derived according to
\begin{eqnarray}\label{camposEHp}
\begin{array}{cc}
\textbf{E}_p(\textbf{x})= i k_0 [\frac{1}{k_1^2} \frac{\partial }{\partial z}\frac{\partial a_p}{\partial x}\hat{x}+
\frac{1}{k_1^2} \frac{\partial }{\partial z} \frac{\partial a_p}{\partial y}\hat{y}+
(1+\frac{1}{k_1^2}\frac{\partial^2 }{\partial z^2}) a_p \hat{z}],\\ 
\textbf{H}_p(\textbf{x})=  
\frac{\partial a_p}{\partial y} \hat{x}
-\frac{\partial a_p}{\partial x} \hat{y}
\end{array}
\end{eqnarray} 
\begin{eqnarray}\label{camposEHs}
\begin{array}{cc}
\textbf{E}_s(\textbf{x})= 
-\frac{\partial a_s}{\partial y}\hat{x}
+\frac{\partial a_s}{\partial x}\hat{y}, \\
\textbf{H}_s(\textbf{x})= i k_0 \varepsilon_1  [
\frac{1}{k_1^2} \frac{\partial }{\partial z}\frac{\partial a_s}{\partial x} \hat{x}+
\frac{1}{k_1^2} \frac{\partial }{\partial z} \frac{\partial a_s}{\partial y}\hat{y}
(1+\frac{1}{k_1^2}\frac{\partial^2}{\partial z^2}) a_s \hat{z} 
]
\end{array}
\end{eqnarray} 
%
The Fourier representation of scalar potentials $a_p(\textbf{x})$ and $a_s(\textbf{x})$ 
is 
\begin{eqnarray}\label{potencialcapam}
\begin{array}{l}
a_\tau^{(m)}(\textbf{x})=\frac{k_0}{2 \pi} \int_{-\infty}^{\infty} d\alpha\,d\beta \,f^{(m)}_\tau(\alpha,\beta,z,z_D) e^{i(\alpha x+\beta y)}
\end{array}
\end{eqnarray}
where functions  $f_\tau^{(m)}(\alpha,\beta,z,z_D)$ ($m=1,\,2,\,3$) depend on the location of the source and of the polarization mode $\tau=p,s$. 
The integrand in (\ref{potencialcapam}) is written as 
\begin{eqnarray}\label{fmedio1}
\begin{array}{l}
f_\tau^{(1)}(\alpha,\beta,z,z_D)=\frac{1}{\gamma^{(1)}} d_\tau \,e^{i\gamma^{(1)}|z-z_D|}+A^{(1)}_\tau\,d_\tau\,e^{i\gamma^{(1)}z}, 
\end{array}
\end{eqnarray}
%
\begin{eqnarray}\label{fmedio2}
\begin{array}{l}
f_\tau^{(2)}(\alpha,\beta,z,z_D)= A^{(2)}_\tau\,d_\tau\,e^{i\gamma^{(2)}z}+B^{(2)}_\tau\,d_\tau\,e^{-i\gamma^{(2)}z},
\end{array}
\end{eqnarray}
%
\begin{eqnarray}\label{fmedio3}
\begin{array}{l}
f_\tau^{(3)}(\alpha,\beta,z,z_D)= B^{(3)}_\tau\,d_\tau\,e^{-i\gamma^{(1)}z},
\end{array}
\end{eqnarray}
%
where the superscript $m=1,\,2,\,3$ denotes medium 1 ($z>d$), medium 2 ($0<z<d$) or medium 3 ($z<0$), and $\gamma^{(j)}=\sqrt{k_j^2-(\alpha^2+\beta^2)}$, with $k_j^2=k_0^2\varepsilon_{j}$ ($j=1,\,2$), is the normal component of the wave vector in each homogeneous region, $k_0=\omega/c$ is the modulus of the photon wave vector in vacuum, $\omega$ is the angular frequency, $c$ is the vacuum speed of light. The former term in Eq. (\ref{fmedio1}) corresponds to the direct field emitted by the dipole placed at $\textbf{x}=\textbf{x}_D=z_D \hat{z}$ \cite{cuevas0,novotny} and
%
%
the spectral functions $d_\tau$ are  given by \cite{cuevas0,novotny}  
\begin{eqnarray}\label{pym}
\begin{array}{l}
d_s=\frac{k_1^2}{k_0 \varepsilon} \left[-\frac{\beta}{\alpha^2+\beta^2}p_x+\frac{\alpha}{\alpha^2+\beta^2}p_y\right] 
\\  
d_p^{\pm}=\mp\frac{\alpha \gamma^{(1)}}{\alpha^2+\beta^2}p_x\mp\frac{\beta \gamma^{(1)}}{\alpha^2+\beta^2}p_y+p_z.  
\end{array}
\end{eqnarray}
The complex coefficients $A^{(m)}_\tau$ and $B^{(m)}_\tau$  in Eqs. (\ref{fmedio1}) to (\ref{fmedio3})  correspond to   the amplitude of upgoing ($+z$ propagation direction) and downgoing ($-z$ propagation direction) plane waves, respectively, and they are  solutions of Helmholtz equation. 
There are two types of boundary conditions which must fulfill the solutions given by Eqs. (\ref{potencialcapam}) to (\ref{fmedio3}), boundary conditions at $z=\pm \infty$ and boundary conditions at interfaces $z=0$ and $z=d$. The former requires either outgoing waves at infinity  or exponentially decaying waves at infinity, depending on the values of $\alpha$, $\beta$ and $\omega$. 
The boundary conditions on interfaces   $z=0$ and $z=d$ impose that
%
\begin{eqnarray} 
\frac{1}{\varepsilon_{m}} \frac{\partial a_p^{(m)}}{\partial z}|_{z=d_{m}}=\frac{1}{\varepsilon_{m+1}} \frac{\partial a_p^{(m+1)}}{\partial z}|_{z=d_{m}}, \label{cc2p} \nonumber\\
a_p^{(m)}|_{z=d_{m}}-a_p^{(m+1)}|_{z=d_{m}}= 
\frac{4\pi  \sigma}{c} \frac{i}{k_0 \varepsilon_{m+1}} \frac{\partial a_p^{(m)}}{\partial z}|_{z=d_{m}}   \label{cc1p}
\end{eqnarray}
for $p$ polarization, and
%
\begin{eqnarray}
a_s^{(m)}|_{z=d_{m}}=a_s^{(m+1)}|_{z=d_{m}},   \label{cc2s} \nonumber\\
\frac{1}{\mu_{m}} \frac{\partial a_s^{(m)}}{\partial z}|_{z=d_{m}}-\frac{1}{\mu_{m+1}} \frac{\partial a_s^{(m+1)}}{\partial z}|_{z=d_{m}}=
-\frac{4\pi}{c} i k_0 a_s^{(m)}|_{z=d_{m}} \label{cc1s}
\end{eqnarray}
for $s$ polarization, where $\sigma$ is the graphene conductivity, $d_1=d$ and $d_2=0$.
To obtain the complex amplitudes  $A^{(m)}_\tau$ and $B^{(m)}_\tau$  
we must combine     Eq. (\ref{potencialcapam}), with $f_\tau^{(m)}$ given by Eqs. (\ref{fmedio1}) to (\ref{fmedio3}),  with conditions (\ref{cc2p}) and (\ref{cc2s})  for $\tau=p$ and  $\tau=s$ polarization, respectively. 

The amplitudes corresponding to region outside the waveguide (region with $m=1$ and $m=3$) are
\begin{eqnarray} \label{a1}
A^{(1)}_\tau=\frac{1}{\gamma^{1}}   
r_\tau^{(1,3)} e^{i\gamma^{(1)}(z_D-2d)}, 
\end{eqnarray}  
\begin{eqnarray} \label{a3}
B^{(3)}_\tau=\frac{1}{\gamma^{1}} t_\tau^{(1,3)} e^{i\gamma^{(1)}(z_D-d)},
\end{eqnarray}  
where 
\begin{eqnarray} \label{a1b}
r_\tau^{(1,3)}=   
\frac{r^{(1,2)}_\tau +r^{(2,1)}_\tau F_\tau e^{i \gamma^{(2)} 2 d} }{1-(r^{(2,1)}_\tau)^2  e^{i \gamma^{(2)} 2 d} },  
\end{eqnarray} 
\begin{eqnarray} \label{a2b}
t_\tau^{(1,3)}=   
\frac{t^{(1,2)}_\tau t^{(2,1)}_\tau e^{i \gamma^{(2)} d} }{1-(r^{(2,1)}_\tau)^2 e^{i \gamma^{(2)} 2 d} },  
\end{eqnarray} 
and
\begin{eqnarray}\label{F} 
F_\tau=t^{(1,2)}_\tau t^{(2,1)}_\tau-r^{(1,2)}_\tau r^{(2,1)}_\tau. 
\end{eqnarray}  
The complex amplitudes
\begin{eqnarray}\label{fresnelp}
r^{(i,j)}_p=\frac{\frac{\gamma^{(i)}}{\varepsilon_i}-\frac{\gamma^{(j)}}{\varepsilon_j}+\frac{4 \pi \sigma}{c k_0} \frac{\gamma^{(i)}}{\varepsilon_i} \frac{\gamma^{(j)}}{\varepsilon_j} }{\frac{\gamma^{(i)}}{\varepsilon_i}+\frac{\gamma^{(j)}}{\varepsilon_j}+\frac{4 \pi \sigma}{c k_0} \frac{\gamma^{(i)}}{\varepsilon_i} \frac{\gamma^{(j)}}{\varepsilon_j}},\\
t^{(i,j)}_p=\frac{2 \frac{\gamma^{(i)}}{\varepsilon_i} }{\frac{\gamma^{(i)}}{\varepsilon_i}+\frac{\gamma^{(j)}}{\varepsilon_j}+\frac{4 \pi \sigma}{c k_0} \frac{\gamma^{(i)}}{\varepsilon_i} \frac{\gamma^{(j)}}{\varepsilon_j}},
\end{eqnarray}
are the Fresnel reflection and transmission coefficients, respectively, for $p$ polarization, whereas 
\begin{eqnarray}\label{fresnels}
r^{(i,j)}_s=\frac{\frac{\gamma^{(i)}}{\mu_i}-\frac{\gamma^{(j)}}{\mu_j}-\frac{4 \pi k_0 \sigma}{c}  }{\frac{\gamma^{(i)}}{\mu_i}+\frac{\gamma^{(j)}}{\mu_j}+\frac{4 \pi \sigma k_0}{c} },
\\
t^{(i,j)}_s=\frac{2 \frac{\gamma^{(i)}}{\mu_i} }{\frac{\gamma^{(i)}}{\mu_i}+\frac{\gamma^{(j)}}{\mu_j}+\frac{4 \pi \sigma k_0}{c} },
\end{eqnarray}
are the Fresnel reflection and transmission coefficients, respectively, for $s$ polarization.

The potential of the scattered field in the  medium $m=1$ can be obtained subtracting the first term in Eq. (\ref{fmedio1}) corresponding  to the primary dipole field, 
\begin{eqnarray}\label{fsmedio1}
\begin{array}{l}
f_\tau^{(1)}(\alpha,\beta,z,z_D)|_{\,scatt}=f_\tau^{(1)}(\alpha,\beta,z,z_D)-\\
\frac{1}{\gamma^{(1)}} d_\tau \,e^{i\gamma^{(1)|z-z_D|}}.
\end{array}
\end{eqnarray}
%

Introducing Eq. (\ref{fsmedio1}) into Eq. (\ref{potencialcapam}), and using Eqs. (\ref{camposEHp}) and (\ref{camposEHs}) we obtain an expression for the scattered electric field on region $z>d$ 
\begin{eqnarray}\label{campoE1}
\textbf{E}^{(1)}(\textbf{x})|_{scatt}= \frac{i k_0^2}{2\pi k_1^2}\int_{-\infty}^{+\infty} \Bigg\{ \left[
-\alpha \gamma^{(1)}\hat{x}
-\beta \gamma^{(1)}\hat{y}
+(\alpha^2+\beta^2)\hat{z} 
 \right]\nonumber \\ 
\times A^{(1)}_p\,d_p^{-}  + 
\frac{k_1^2}{ k_0}\left[
-\beta \hat{x}
+\alpha \hat{y}
 \right] A^{(1)}_s\,d_s^{-} \Bigg\} 
e^{i \gamma^{(1)}z} e^{i[\alpha x+\beta y]} d\alpha d\beta, 
\end{eqnarray}
where $A^{(1)}_p$ and  $A^{(1)}_s$ are given by Eq. (\ref{a1}). In a similar way, we obtain an expression for the electric field on region $z<0$ 
\begin{eqnarray}\label{campoE3}
\textbf{E}^{(3)}(\textbf{x})= \frac{i k_0^2}{2\pi k_1^2}\int_{-\infty}^{+\infty} \Bigg\{ \left[
-\alpha \gamma^{(1)}\hat{x}
-\beta \gamma^{(1)}\hat{y}
+(\alpha^2+\beta^2)\hat{z} 
 \right] B^{(3)}_p\,d_p^{-} \nonumber \\ + 
\frac{k_1^2}{ k_0}\left[
-\beta \hat{x}
+\alpha \hat{y}
 \right] B^{(3)}_s\,d_s^{-} \Bigg\}
\times e^{-i \gamma^{(1)}z} e^{i[\alpha x+\beta y]} d\alpha d\beta, 
\end{eqnarray}
where $B^{(3)}_p$ and  $B^{(3)}_s$ are given by Eq. (\ref{a3}).

\section{Graphene conductivity} \label{grafeno}
\setcounter{equation}{0}
\renewcommand{\theequation}{C{\arabic{equation}}}
We consider the  graphene layer as an infinitesimally thin, local and isotropic two--sided layer with frequency--dependent surface conductivity $\sigma(\omega)$ given by the Kubo formula \cite{falko,milkhailov}, which can be read as  $\sigma= \sigma^{intra}+\sigma^{inter}$, with the intraband and interband contributions being
\begin{equation} \label{intra}
\sigma^{intra}(\omega)= \frac{2i e^2 k_B T}{\pi \hbar (\omega+i\gamma_c)} \mbox{ln}\left[2 \mbox{cosh}(\mu_c/2 k_B T)\right],
\end{equation}  
\begin{eqnarray} \label{inter}
\sigma^{inter}(\omega)= \frac{e^2}{\hbar} \Bigg\{   \frac{1}{2}+\frac{1}{\pi}\mbox{arctan}\left[(\omega-2\mu_c)/2k_BT\right]-\nonumber\\
\frac{i}{2\pi}\mbox{ln}\left[\frac{(\omega+2\mu_c)^2}{(\omega-2\mu_c)^2+(2k_BT)^2}\right] \Bigg\},
\end{eqnarray}  
where $\mu_c$ is the chemical potential (controlled with the help of a gate voltage), $\gamma_c$ the carriers scattering rate, $e$ the electron charge, $k_B$ the Boltzmann constant and $\hbar$ the reduced Planck constant.


\begin{thebibliography}{100}


\bibitem{barnes} Andrew P. and  Barnes W. L. 2004 
Energy Transfer Across a Metal Film Mediated by Surface Plasmon Polaritons 
{\it Science} {\bf 306} 1002 

\bibitem{arruda1}  Arruda T. J.,  Bachelard R.,   Weiner J.,  Slama S., and  Courteille P. W. 2017 
Fano resonances and fluorescence enhancement of a dipole emitter near a plasmonic nanoshell 
{\it Phys. Rev.} A {\bf 96} 043869 

\bibitem{pra84}
 Marocico C. A. and  Knoester J. 2011 Effect of surface--plasmon polaritons on spontaneous emission and intermolecular energy--transfer rates in multilayered geometries,''
{\it Phys. Rev.} A {\bf 84}  053824 


\bibitem{shahbazyan}  Pustovit V. N., and  Shahbazyan T. V. 2011 
Resonance energy transfer near metal nanostructures mediated by surface plasmons,''
{\it Phys. Rev.} B {\bf 83} 085427 

\bibitem{maier_nature} 
Tame M S, McEnery K R, Ozdemir K, Lee J,  Maier S A  and   Kim M S 2013 
Quantum plasmonics
{\it Nat. Phys.} {\bf 9} 329--340

\bibitem{wubs}  Wubs M. and  Vos W. L.  2016 
F\"orster resonance energy transfer rate in any dielectric nanophotonic medium with weak dispersion 
{\it New J. Phys.} {\bf 18} 053037 

\bibitem{forster}
 Medintz I.L., Hildebrandt N., Eds. 
{\it FRET--F\"orster Resonance Energy Transfer. From Theory to Applications} 
(Wiley--VCH, Weinheim, Germany 2013)

\bibitem{superradiance}  Huidobro P. A.,  Nikitin A. Y.,  Gonzalez--Ballestero C.,  Mart\'in--Moreno L., and Garc\'ia--Vidal F. J. 2012 Superradiance mediated by graphene surface plasmons 
{\it Phys. Rev.} B {\bf 85} 155438

\bibitem{grafeno3}
 Koppens F. H. L.,  Chang D. E. and  Garc\'ia de Abajo F. J. 2011 
Graphene plasmonics: A platform for strong light--matter interactions 
{\it Nano Lett.} {\bf 11} 3370--3377

\bibitem{QI} Karanikolas V. and Paspalakis E.  2018 Plasmon--Induced Quantum Interference near Carbon Nanostructures
{\it J. Phys. Chem.} C {\bf 122} 14788--14795

\bibitem{jablan}
Jablan J., Soljacic M., Buljan H. 2013 
Plasmons in graphene: fundamental properties and potential applications
{\it Proc. IEEE} {\bf 101} 1689--1704   

\bibitem{rana}
 Rana  F.  2008 
Graphene terahertz plasmon oscillators
{\it IEEE Trans  Nano--technol} {\bf 7} 91--99

\bibitem{Naserpour}
Naserpour M., Zapata--Rodr\'iguez C. J., Vukovic S. M., Pashaeiadl H. and  Belic M. R. 2017 
Tunable invisibility cloaking by using isolated graphene-coated nanowires and dimers 
{\it Sci. Rep.} {\bf 7} 12186

\bibitem{Alu}
Chen P.Y.  and  Alu A. 2011 
Atomically thin surface cloak using graphene monolayers 
{\it ACS Nano} {\bf 5}  5855--5863 

\bibitem{filter}  Filter R., Farhat M., Steglich M., Alaee R.,  Rockstuhl R. and  Lederer F.    2013
Tunable graphene antennas for selective enhancement of THz--emission 
{\it Opt. Exp.} {\bf 21}  3737 

\bibitem{gomez_diaz}  Tamagnone M.,  G\'omez--D\'iaz J. S.,  Mosig J. R. and  Perruisseau--Carrier J. 2012
 Reconfigurable terahertz plasmonic antenna concept using a graphene stack 
{\it Appl. Phys. Lett.} {\bf 101} 214102 

\bibitem{jornet}   Llatser  I.,  Kremers C., Cabellos--Aparicio A., Jornet J. M.,  Alarc\'on E., Chigrin D. N,  Graphene--based nano--patch antenna for terahertz radiation 2012
{\it Photonics and Nanostructures--Fundamentals and Applications} {\bf 10} 353--358 

\bibitem{cuevas6}  Cuevas M. 2018 
Theoretical investigation of the spontaneous emission on graphene plasmonic antenna in THz regime 
{\it Superlattices and Microstructures} {\bf 122} 216--227

\bibitem{marocico}
 Karanikolas V. D.,  Marocico C. A., and  Bradley A. L. 2015 
Dynamical tuning of energy transfer efficiency on a graphene monolayer  
{\it Phys. Rev.} B {\bf 91} 125422

\bibitem{arruda2}  Arruda T. J., Bachelard R., Weiner J.,  Courteille P. W. 2018
 Tunable Fano resonances in the decay rates of a point--like emitter near a graphene-coated nanowire  
{\it arXiv}:1809.05208 

\bibitem{LSP}
Christensen T, Jauho A--P, Wubs M and Mortensen N  2015
Localized plasmons in graphene--coated nanospheres  
{\it Phys. Rev.} B {\bf 91} 125414                            

\bibitem{grafeno2}
 Kort--Kamp W. J. M.,  Amorim B.,  Bastos G.,  Pinheiro F. A.,  Rosa F. S. S.,  Peres N. M. R., and  Farina C. 2015 
Active magneto--optical control of spontaneous emission in graphene  
{\it Phys. Rev.} B {\bf 92} 205415  

\bibitem{cuevas3} Cuevas M. Spontaneous emission in plasmonic graphene subwavelength wires of arbitrary sections 2018   
{\it Journal of Quantitative Spectroscopy and Radiative Transfer}  {\bf 206}  157--162 

\bibitem{cuevas3bis}  Cuevas M. 2017 Graphene coated subwavelength wires: a theoretical investigation of emission and radiation properties 
{\it Journal of Quantitative Spectroscopy and Radiative Transfer} {\bf 200} 190--197

\bibitem{cuevas4}  Cuevas M. 2018 Enhancement, suppression of the emission and the energy transfer by using a graphene subwavelength wire
{\it Journal of Quantitative Spectroscopy and Radiative Transfer}  {\bf 214} 8--17



\bibitem{biehs} Biehs S.A., and  Agarwal G. S. 2013 Large enhancement of Förster resonance energy transfer on graphene platforms 
{\it Appl. Phys. Lett.} {\bf 103} 243112 

\bibitem{shahbazyan2} 
 Velizhanin K. A., Shahbazyan T. V. 2012 Long--range plasmon-assisted energy transfer over doped graphene 
{\it Phys. Rev.} B {\bf 86} 245432  
 


\bibitem{wire}   Hua X. M., and  Gersten J. I. 1989 Enhanced energy transfer between donor and acceptor molecules near a long wire or fiber  
{\it The Journal of Chemical Physics} {\bf 91}  1279 









\bibitem{nikitin}  Nikitin A. Yu.,  Guinea F.,  Garcia--Vidal F. J., and  Martin--Moreno L. 2011 
Fields radiated by a nanoemitter in a graphene sheet  
{\it Phys. Rev.} B {\bf 84}   195446 

\bibitem{cuevas0}  Cuevas M. 2016 
Surface plasmon enhancement of spontaneous emission in graphene waveguides  
{\it J. Opt.} {\bf 18} 105003 

\bibitem{novotny}
Novotny L, and Hecht B  
{\it Principles of Nano--Optics}, (Cambridge University Press: New
York, 2006).

\bibitem{abramowitz}  Abramowitz M. and   Stegun I. A.,  {\it Handbook of Mathematical Functions}  
(Dover New York, 1965) 



\bibitem{wait}  Wait J. R.   1967 Asymptotic Theory for Dipole Radiation in the Presence of a Lossy Slab Lying on a Conducting Half-space  
{\it IEEE Transactions on Antennas and Propagation}  {\bf 15} 645--648

\bibitem{michalski} Michalski K.A. and  Mosig J.R. 2016 
The Sommerfeld half--space problem revisited: from radio frequencies and Zenneck waves to visible light and Fano modes                                                                     
{\it Journal of Electromagnetic Waves and Applications} {\bf 30} 1--42 

\bibitem{paulus}   Paulus M.,  Gay--Balmaz P., and  Martin O. J. F. 2000
   Accurate and efficient computation of the Green's tensor for stratified media,''  
{\it Phys. Rev.} E {\bf 62} 5797--5807  



 





\bibitem{falko}  
Falkovsky FA 2008 
Optical properties of graphene and IV--VI semiconductors 
{\it Phys. Usp.} {\bf 51} 887--897 

\bibitem{milkhailov} 
Milkhailov SA and Siegler K (2007)
New electromagnetic mode in graphene 
{\it Phys. Rev. Lett.} {\bf 99}  016803 



\end{thebibliography}
\end{document}